\documentclass[a4paper,english, 12pt]{article}
\pdfoutput=1
\usepackage[T1]{fontenc}
\usepackage{bbold}
\usepackage{graphicx}
\usepackage{bbm}
\usepackage{color}
\usepackage{amsmath}
\usepackage{amssymb}
\usepackage{amsfonts}

\newcommand{\Ef}{ \mathcal{E} }
\newcommand{\Af}{ \mathcal{A} }

\newcommand{\Bf}{ \mathcal{B} }

\newcommand{\Id}{ \mathrm{d} }

\newcommand{\Diff}[1]{ D_{#1}}
\newcommand{\VDiff}[1]{ \vec{D}_{\vec{#1}}}

\begin{document}

\title{Simulation of Pair Production \\ in Extreme Strong EM Fields}

\author{ 
\normalsize{Dániel Berényi$^{1, 2}$, Péter Lévai$^2$, Vladimir Skokov$^3$} \\ 
\footnotesize{$^1$Eötvös Loránd University, Budapest, Hungary} \\ 
\footnotesize{$^2$MTA WIGNER RCP, RMKI, Budapest, Hungary} \\
\footnotesize{$^3$Brookhaven National Laboratory, Upton, USA}
}
\date{}
\maketitle
\begin{abstract}
In this article we review a theoretical framework for pair production from strong external electromagnetic fields. We propose a numerical method to solve the resulting equations of motion and present results for both cases of spatially homogeneous and inhomogeneous electric fields.

\end{abstract}

\maketitle

\section{Introduction}

Since the prediction of $e^+ e^-$ pair production from vacuum in a strong electric field by J. Schwinger \cite{Schwinger}, extensive theoretical research has been focused on the non-perturbative QED processes. While earlier, the experimental observation was way beyond feasible, recent advances in laser technology outperform the expectations. With the advent of Chirped Pulse Amplification \cite{CPA} and the creation of femtosecond and attosecond intense laserpulses, the threshold of pair production is being approached by the extreme field laboratories\cite{ELI}.

At the same time, the theoretical description of the phenomena has been also developed. For a long time, different simple, but analitically solvable models of the external field were investigated but it was evident, that the full description of a realistic field configuration is well beyond the applicability of those models \cite{Popov}. Recently, a Wigner function-based formalism was proposed, that can describe arbitrary field configurations and can describe a self-consistent description of the time evolution of the particle --- anti-particle quantum system \cite{Rafelski, Skokov, Hebenstreit}.

Most investigations were performed in the spatially homogeneous limit, because the calculations were beyond the capabilities of the available computers. With the advent of cheap computing devices, the calculations in multiple dimensions and with spatial inhomogenities has become feasible. We present a numerical method that is ideal for the hardware and can efficiently solve the complex differential equations of pair production.

First, we briefly review the theoretical framework; next, we discuss the pair yield in chirped homogeneous electric field; and finally, present first results for an inhomogeneous field.

\section{The Dirac-Heisenberg-Wigner formalism}

In a classical system, the one particle distribution function is used to characterize the time evolution. In the quantum case the Wigner function ($W$) plays a similar role  \cite{Wigner}. It is defined from the following density operator ($\hat{C}$):

\begin{equation}
	 W(\vec{x}, \vec{p}, t) = -\frac{1}{2} \int e^{-i\vec{p}\vec{s}} \langle 0|\hat{C}(\vec{x}, \vec{s}, t)|0\rangle \Id ^3 s,
\end{equation}

where

\begin{equation}
	\hat{C}(\vec{x}, \vec{s}, t) = \exp\left[-ie\int_{-1/2}^{1/2} \vec{\Af}(\vec{x} + \lambda \vec{s}, t) \vec{s} \Id \lambda\right] \left[\Psi(\vec{x}+\frac{\vec{s}}{2}, t), \bar{\Psi}(\vec{x}-\frac{\vec{s}}{2}, t)\right].
\end{equation}

Here $\vec{\Af}$ is the vector potential of the electromagnetic field. An equation of motion for the Wigner function can be derived by applying the time derivate to $W$. The derivation can be found elsewhere \cite{Rafelski, Hebenstreit}.  The equation of motion reads:

\begin{align}
 \Diff{t}W = -\frac{1}{2}\VDiff{x} \left[ \gamma^0 \vec{\gamma}, W\right]-im\left[ \gamma^0, W\right]-iP\left\{\gamma^0 \vec{\gamma}, W\right\}
\end{align}

where the following non-local operators are introduced:

\begin{equation}
\label{nonloc1}
D_t =
 \partial_t + e\vec{\Ef}(\vec{x},t) \vec{\nabla}_{\vec{x}} -\frac{e\hbar^2}{12}(\vec{\nabla}_{\vec{x}} \vec{\nabla}_{\vec{p}})^2 \vec{\Ef}(\vec{x},t) \vec{\nabla}_{\vec{p}} + \ldots ;
\end{equation}
\begin{equation}
\VDiff{x} =
\vec{\nabla}_{\vec{x}} + e\vec{\Bf}(\vec{x},t) \times \vec{\nabla}_{\vec{x}} -\frac{e\hbar^2}{12}(\vec{\nabla}_{\vec{x}} \vec{\nabla}_{\vec{p}})^2 \vec{\Bf}(\vec{x},t) \times \vec{\nabla}_{\vec{p}} + \ldots ;
\end{equation}
\begin{equation}
\label{nonloc3}
\vec{P} =
\vec{p} + \frac{e\hbar}{12}(\vec{\nabla}_{\vec{x}} \vec{\nabla}_{\vec{p}})\vec{\Bf}(\vec{x},t) \times \vec{\nabla}_{\vec{p}} + \ldots .
\end{equation}

\begin{figure}
		\centering
		\includegraphics[width=0.87\textwidth]{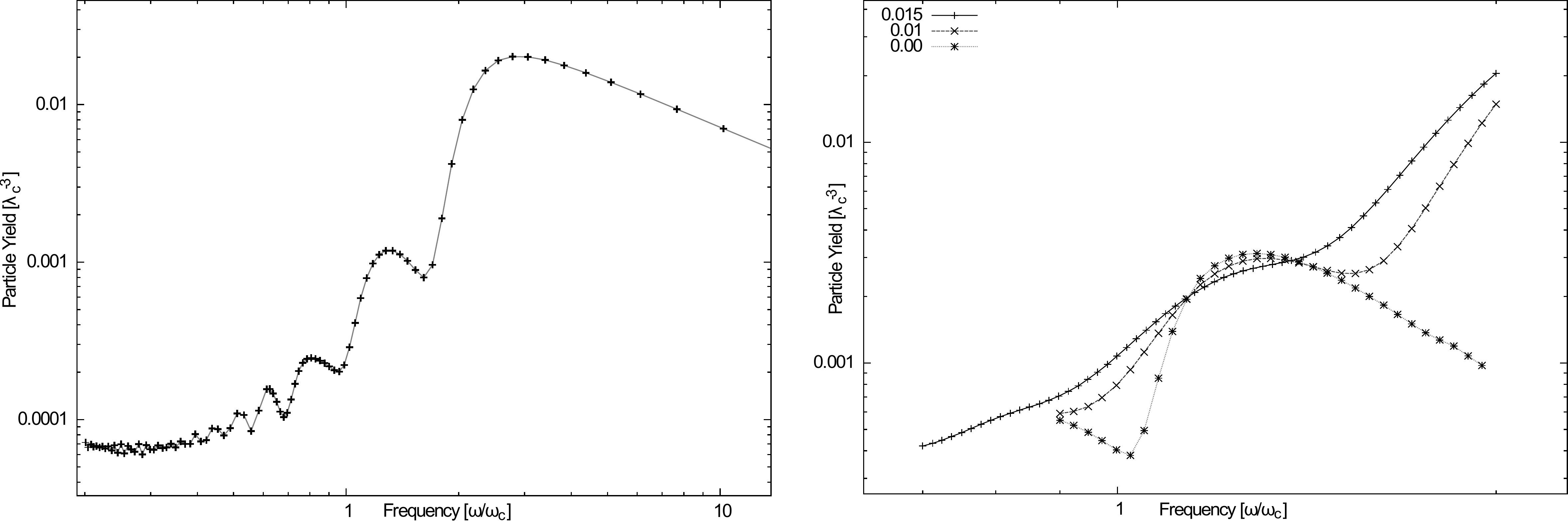}
\caption{Resonant transition from adiabatic tunneling to anti-adiabatic multi photon pair creation (left). Zoom in around the critical frequency ($\omega_c$) for different chirp parameters (right).}
\end{figure}

with $\vec{\Ef}$ and $\vec{\Bf}$ denoting the electric and magnetic field respectively.

In the following, we will use the expansion of the Wigner function: $W(\vec{x}, \vec{p}, t) = 1/4[\mathbbm{1s}+i\gamma_5\mathbbm{p}+\gamma^\mu \mathbbm{v}_\mu+\gamma^\mu \gamma_5 \mathbbm{a}_\mu + \sigma^{\mu\nu}\mathbbm{t}_{\mu\nu}]$. In this basis, the particle density can be calculated as:

\begin{equation}
f(\vec{x}, t) = 2 + \frac{\mathbbm{s}m + \vec{\mathbbm{v}}\vec{p}}{\sqrt{m^2+\vec{p}^2}}
\end{equation}

In the limit of $\vec{\Bf} = 0$ and $\vec{\Ef}(\vec{x}, t) = \Ef(t)$, one restores the well-known quantum kinetic equation \cite{Skokov}:

\begin{eqnarray}
\label{eq:qk}
	\frac{\Id f}{\Id t} & = & v e\Ef \sqrt{m^2 + \vec{p}^2_{\perp}} / (m^2+\vec{p}^2), \\
	\frac{\Id v}{\Id t} & = & \frac{1}{2} (1-2f) e\Ef \sqrt{m^2 + \vec{p}^2_{\perp}} / (m^2+\vec{p}^2)  - 2 u \sqrt{m^2+\vec{p}^2}, \\
	\frac{\Id u}{\Id t} & = & 2 v \sqrt{m^2+\vec{p}^2};
\end{eqnarray}

where $f$ is the particle density ($u$ and $v$ are auxiliary functions) and the kinetic and canonic momenta has the usual connection: $\vec{p} = (\vec{q}_{\perp}, q_{\parallel} - e\Af(t))$. While this equation was known long ago, the total pair production-dependence on field parameters was not discussed for complex time-dependence of the external field.

\section{Numerical Results}
To get as close to the experimental laser fields as possible in the quantum kinetic model we consider the following two models for the electric fields: a packet like $\Ef(t) = \Ef_0 \exp[-\left(t / \tau\right)^2] \cos \left( \phi + \omega t\right)$ and a similar, but chirped field: $\Ef(t) = \Ef_0 \exp[-\left(t / \tau\right)^2] \cos \left( \phi + \omega t + c t^2\right)$ where the field strength is measured in $\Ef_c = \frac{m^2 c^3}{e \hbar}$ units, also we set $\hbar = c = 1$. The system exhibits a resonant transition at the frequency $\omega_c = \frac{e \Ef}{mc}$ between two qualitatively different regions: tunneling and multi photon pair creation. Figure 1 has been calculated for the packet-like field with $\Ef_0 = 0.5 \cdot \Ef_{c}$, $\phi = 0$ and $\tau$ chosen in such a way as to have 5 cycles in the $0.1 \cdot \Ef_{max}$ envelope. We investigated the effect of adding chirp to the field (Figure 1., right panel). An increase of the chirp results in the widening of the peaks and the rise of the average yield. If experiments aim to investigate this region for resonances, the chirp may become a natural control parameter.

\begin{figure}
		\centering
     \includegraphics[width=0.87\textwidth]{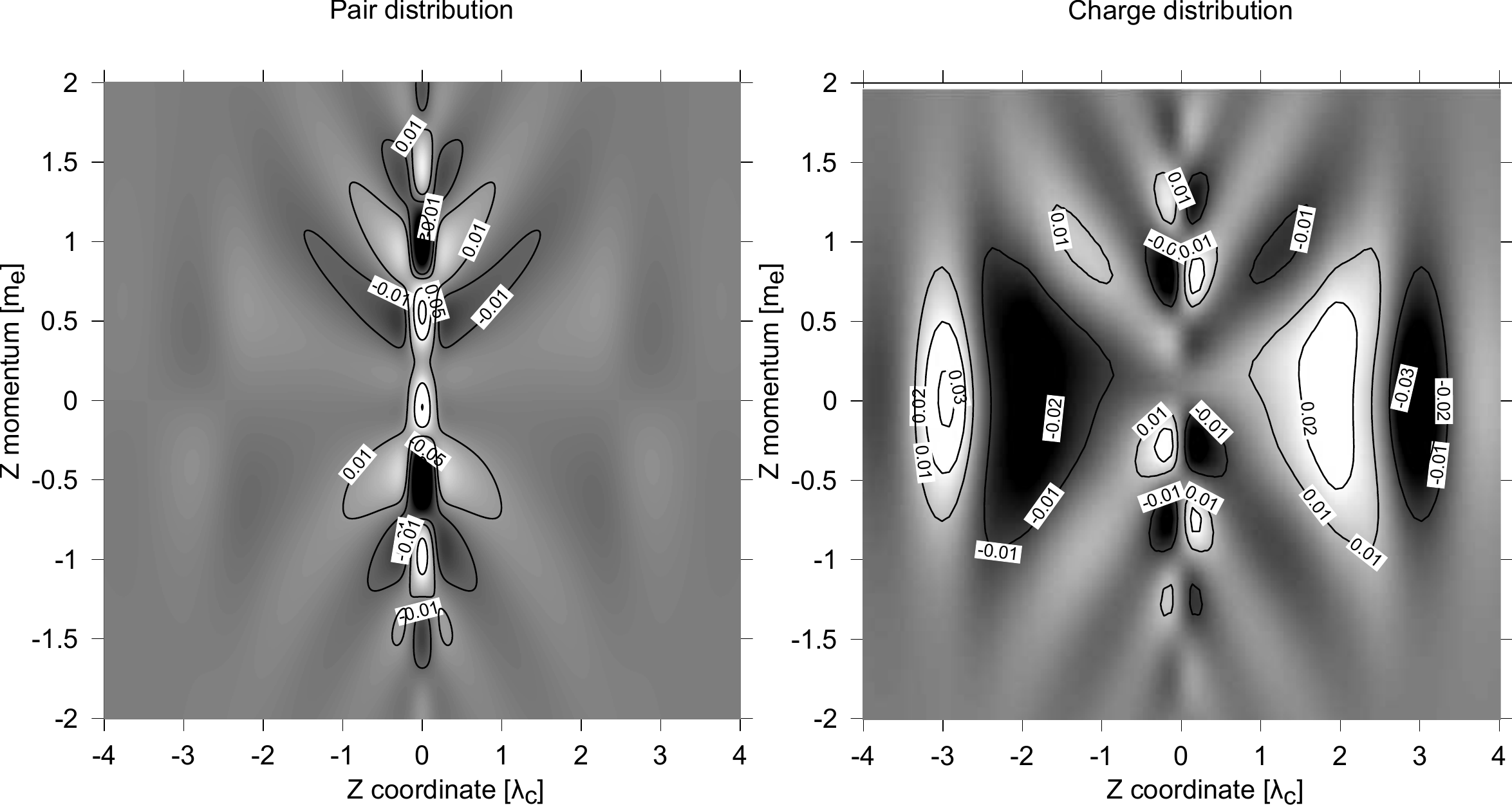}
     \caption{Pair density (left) and Charge density (right). The density scale is arbitrary.}
\end{figure}

In the spatial inhomogeneous case, no simplifications are possible, and we must use all 16 components of the Wigner function. Problem arises because of the high dimensionality and the derivative expansion of the non-local operators (\ref{nonloc1})-(\ref{nonloc3}). We successfully used spectral expansion to solve the system of equations in 1+1 dimensions at the leading order of the derivative expansion. The method can be easily generalized to higher dimensions and can reliably resolve higher order terms in the derivative expansion. Details of the method will be discussed in a forthcoming paper. Here we illustrate the results for the following simple field: $\Ef(z, t) = \Ef_0 \exp{\left[-\left(z^2 / \sigma^2 + t^2 / \tau^2\right)\right]}$. Figure 2 show the pair density and the charge density of the system after the pulse. The spatial inhomogenity creates interesting structures in the distributions. With such methods, it becomes possible to understand the microscopic spatiotemporal evolution of the system in a comprehensible way.

\section{Conclusion}
We reviewed the Wigner formalism of pair production for homogeneous and inhomogeneous electric fields. For the homogeneous case we investigated the effect of chirp on the produced number of pairs. We found that the chirp parameter tunes the contrast of the resonance peaks.

We developed a method based on spectral expansion to solve the Wigner equation in 1+1 dimensions. This method can be used to efficiently extend the analysis to the spatial structure of the quantum distributions characterising the pair production. This opens new possibilities to study more realistic laser fields and to provide more adequate predictions.

This work was supported by the Hungarian OTKA Grant NK 77816.


\end{document}